\documentclass[journal, onecolumn]{IEEEtran}
\ifCLASSINFOpdf
  \usepackage[pdftex]{graphicx}
\else
\fi
\usepackage[cmex10]{amsmath}
\begin{document}
%
\title{Leave-One-Out Learning with Log-Loss}
%
%
%
\newtheorem{theorem}{Theorem}
\newtheorem{corollary}{Corollary}
\newtheorem{lemma}{Lemma}


\author{Yaniv~Fogel~and~Meir~Feder

\thanks{Meir Feder and Yaniv Fogel are with the School
of Electrical Engineering, Tel Aviv University, Israel}
}

\maketitle

\begin{abstract}

  We study batch learning with log-loss in the \emph{individual} setting, where the outcome sequence is deterministic. Because empirical statistics are not directly applicable in this regime, obtaining regret guarantees for batch learning has long posed a fundamental challenge~\cite{grunwald2011bounds}. We propose a natural criterion based on \emph{leave-one-out regret} and analyze its minimax value for several hypothesis classes. For the multinomial simplex over $m$ symbols, we show that the minimax regret is $\frac{m-1}{N} + o\!\left(\frac{1}{N}\right)$, and compare it to the stochastic realizable case where it is $\frac{m-1}{2N} + o\!\left(\frac{1}{N}\right)$. More generally, we prove that every hypothesis class of VC dimension $d$ is learnable in the individual batch-learning problem, with regret at most $\frac{d\log(N)}{N} + o\!\left(\frac{\log(N)}{N}\right)$, and we establish matching lower bounds for certain classes. We further derive additional upper bounds that depend on structural properties of the hypothesis class. These results establish, for the first time, that universal batch learning with log-loss is possible in the individual setting.

\end{abstract}


%
\IEEEpeerreviewmaketitle

%
%
%
%

 

\section{Introduction}
Information theory suggests an approach to machine learning based on the following assumptions. First, the loss function is the log-loss; that is, a learner provides a probabilistic prediction for the outcome $q(\cdot)$ and attains a loss that is $-\log q(y)$ where $y$ is the resulting outcome. This loss has been justified both theoretically, e.g., \cite{painsky2018universality} and elsewhere, and practically, as most modern practical learning schemes are trained with this loss. Nevertheless, most of the classical learning theory results assume a bounded loss function with a few notable exceptions, e.g., \cite{grunwald2020fast} and \cite{haddouche2021pac}. Second, there is a hypotheses class $\{p_\theta, \theta\in\Theta\}$ which serves as a reference for the proposed learner. There are also several ways to define how the data is generated. In particular, three settings can be defined: a realizable stochastic setting in which the data is generated by one of the models in the class and the goal is to attain the same performance as if the model is known, a mis-specified, or non-realizable, stochastic setting in which the data comes from a stochastic model that is not necessarily in the class and the goal is to attain the optimal projection of the true model over the class, and finally, an individual setting where the data is deterministic, individual and the goal is to attain the performance of the best model in the class that fits that sequence. 

Usually, the information theoretic approach assumes that the goal is to predict the entire outcome sequence in an online manner. Online learning with log-loss is identical to coding the entire sequence via a lossless code, establishing the equivalence between prediction and coding \cite{universal_prediction}. Yet, a common scenario in machine learning is ``batch learning'', where there is a training set and the learner predicts the possible test outcome. Recently, batch learning was also considered from an information-theoretic point of view in the stochastic setting, e.g., in \cite{fogel2018stochastic} and \cite{vituri2024}. However, the definition of the individual setting for batch learning is challenging, as the performance is measured over a single ``test'' example, which can be arbitrary. This challenge has  been recognized, e.g., in \cite{grunwald2011bounds}, presented as an ``open problem''. Actually, due to the difficulty in defining a plausible criterion for the individual setting, the paper suggested to shift the focus to the stochastic, realizable setting.

In recent works, we have suggested two criteria to handle this case, see \cite{fogel2019individual} and \cite{fogel2022another}. Those criteria led to previously known learners that can be thought of as a variation of the celebrated normalized maximum-likelihood (NML) learner. These variations were originally presented in \cite{roos2008bayesian}, \cite{roos2008sequentially}, and \cite{Grunwald2007} as approximations from the NML that can be computed more efficiently. These learners can also be computed for hypothesis classes such a linear regression, and their algorithmic variations also proved to be useful in practice - see \cite{bibas2019deep} \cite{li2021mural} and \cite{fu2021offline} for examples. Nevertheless, as we show in section \ref{pNMLs}, the resulting learners might fail to learn some simple hypothesis classes that are learnable in the stochastic setting. Thus, the criterion for a universal learner in the individual setting does not have a satisfactory and accepted definition. 

It seems that an intuitively appealing and plausible criterion, that may catch the empirical behavior of the data, can be based on defining the regret via a leave-one-out averaging over the sequence. In this paper, we will elaborate on this criterion definition and study its min-max value and optimal solution in several examples. 

The first example we discuss is the multinomial distribution, with no data features, and the models are of i.i.d distributions. In this case, we show that the min-max optimal regret is $\frac{m-1}{N} + o(\frac{1}{N})$, 
and compare it to the stochastic case where $\frac{m-1}{2N} + o(\frac{1}{N})$ can be achieved. 

We also consider hypothesis classes where $x$ is first assigned to one of $2$ partitions, and there is a probability assignment of $y$ that depends only on this partition. We show that when the VC-dimension of the allowable partitions of the data features $x$ is $d$, a regret of $\frac{d\log(N)}{N} + o \left( \frac{d\log(N)}{N} \right)$ is achievable under rather mild conditions. We also show that for every $d$, there is a hypothesis class of VC-dimension $d$ for which the regret is $\frac{d\log(N)}{N} +o(\frac{\log(N)}{N})$. We also consider the case where the probability assignments are deterministic, where we show upper bounds over the regret that depend upon properties of the relevant one-inclusion graph.

\section{Previous Works}

\subsection{Universal Prediction}

Before we survey the batch learning scenario, we will briefly review relevant results from the field of universal prediction, see \cite{universal_prediction}. In the prediction problem, there are no data features $x$'s, and one is only interested in predicting the outcomes $y$'s. Thus, for the log-loss function, there is an equivalence between the problem of assigning a probability to the whole sequence $y^N$ and sequentially assigning a probability for $y_{t+1}$ given $y^t = y_1, y_2, ..., y_t$. Naturally, when using the log-loss function, the hypothesis class takes a probabilistic form: $\{p_{\theta}(y^N), \theta \in \Theta\}$.

There are several settings for the universal prediction problem. The first is the realizable, stochastic setting, where one assumes that the data is generated by one of the probabilities in the hypothesis class. In this case, the regret is averaged over $y^N$:

\begin{align*}
    R_{stochastic} \left( \theta, q(\cdot) \right) = \frac{1}{N}\sum_{y^N} p_\theta(y^N) \log \left( \frac{p_\theta(y^N)}{q(y^N)} \right).
\end{align*}


A possible criterion for determining the learner is 
the redundancy or the min-max regret: 
\begin{align*}
    R^*_{stochastic} = \min_{q(\cdot)} \max_{\theta} R_{stochastic}  \left( \theta, q(\cdot) \right).
\end{align*} 
The celebrated Redundancy-Capacity theorem, due to \cite{gallager1979source}, \cite{davisson1980source}, and \cite{ryabko1979coding} proves that this min-max problem is equivalent to the max-min average regret
which is equal to the capacity of the channel between $Y^N$ and $\theta$, $ \max_{w(\theta)} I(Y^N; \Theta)$. 
The optimal learner turned out to be a Bayesian mixture with the capacity achieving prior over the hypothesis class. For smooth parametric families, 
the regret behaves as $\frac{k\log(N)}{2N}$ where $\theta \in \mathbf{R}^k$, see \cite{rissanen1984universal}, and Jefferys' prior attains the optimal regret, \cite{clarke1994jeffreys}.

The stochastic setting can be extended to an unrealized or mis-specified setting where the data is generated by some stochastic model that is not necessarily in the class. While this setting has been suggested already in \cite{barron1998}, it was not analyzed in depth until recently by \cite{federpolyanskiy2021}, \cite{PainskyFeder2}. 

The most general setting of universal prediction is the individual setting, where one assumes that the outcomes sequence $y^N$ is an arbitrary, deterministic sequence $y^N$:

\begin{equation*}
    R_{individual}\left( y^N, q(\cdot) \right) =  \log \left( \frac{ \max_{\theta} p_\theta(y^N)}{q(y^N)} \right)
\end{equation*}

For this setting, \cite{shtar1987universal} have shown that the min-max optimal predictor is the Normalized Maximum Likelihood (NML):

\begin{equation*}
    q_{NML}(y^N) = \frac{\max_{\theta} p_\theta(y^N)}{ \sum_{\Tilde{y}^N}\max_{\theta} p_\theta(\Tilde{y}^N)}.
\end{equation*}

Interestingly, the proof that the NML is the min-max optimal probability assignment relies on the observation that it attains an equal regret for every $y^N$, which is equal to $\log \left( \sum_{\Tilde{y}^N}\max_{\theta} p_\theta(\Tilde{y}^N) \right)$.

\subsection{Batch Learning}

Consider now the case where a training set $z^{N-1} = (x, y)^{N-1}$ is given, and one is only required to predict the next outcome $y_N$ given its respective data feature $x_N$, with a log-loss.
For the stochastic setting, a natural way to define the regret is as follows:
\begin{align*}
\begin{split}
    R_{stochastic} \left( \theta, q(\cdot) \right) 
    = 
    \frac{1}{N}\sum_{y^N x^N} p(x^N) \cdot p_\theta(y^N|x^N) \log \left( \frac{p_\theta(y_N|y^{N-1}, x^N)}{q(y_N|y^{N-1}, x^N)} \right)
\end{split}
\end{align*}
where again, it is shown in \cite{fogel2018stochastic} that the min-max optimal redundancy is equivalent to a max-min problem whose solution is the conditional capacity of the channel between $y_N$ and $\Theta$ given $z^{N-1}, x_N$, that is $\max_{w(\theta)} I(Y_N; \Theta | Y^{N-1}, X^N)$, which is upper-bounded by $\max_{w(\theta)}\frac{1}{N}I(Y^N;\Theta|X^N)$ for hypothesis classes of the form $p_{\theta}(y^N|x^N) = \prod_{t=1}^N p_\theta(y_t|x_t)$.

For the min-max batch learning problem, there are several results for the multinomial distribution, where there are no data features, $y \in \{0, 1, .., m-1 \}$, $\theta \in \left[0, 1 \right] ^ m $ is the simplex and $p_{\theta}(y^N) = \prod_{t=1}^N \theta(y_t)$. First, \cite{krichevskiy1998laplace} showed that the regret is lower bounded by $\frac{m-1}{2N} + o(\frac{1}{N})$, and that the class of add-$\beta$ learners can only achieve a slightly higher regret of $\frac{0.5092(m-1)}{N} + o(\frac{1}{N})$ . Later on, \cite{braess2004bernstein} have shown that one can achieve a regret of order $\frac{m-1}{2N}$ with an add-$\frac{3}{4}$ rule with some adjustments at the edges. Interestingly, \cite{Komaki2012AsymptoticallyMinimax} have shown that when one does not take the edges into account, the optimal probability assignment becomes add-$1+\sqrt{\frac{1}{6}}$, while Jefferys prior translates to add-$\frac{1}{2}$. 

Batch learning in the stochastic case, the simplest setting, still has some open issues even for the multinomial case: Currently, there is no closed-form expression of the capacity achieving prior or the induced optimal probability assignment.
For more general hypothesis classes, \cite{Forster2002RelativeExpectedInstantaneous} have considered several specific hypothesis classes including Gaussian densities with known variance and unknown mean. 


Batch learning in the individual setting is more challenging. 
A possible reason is that it is not obvious how to construct the proper criterion that utilizes the empirical behavior of the data while dealing with a single test example. Some discussion on this difficulty can be found in \cite{fogel2019individual}. Nevertheless, one plausible and intuitive approach is to utilize a leave-one-out approach. This is the topic of this paper. The exact formulation of the criterion is given below in section 4 and the paper is dedicated to its investigation.

The leave-one-out approach has been discussed and analyzed extensively, but not for the prediction problem with log-loss.
Most previous works have analyzed other loss functions, especially the $0-1$ loss function. In this case, one notable work is \cite{predicting_01_on_randomly_drawn_points}, where a bound over the individual regret of leave-one-out with the $0-1$ was achieved. This was done using an analysis of the \textbf{one inclusion graph}, a graph where the nodes are sequences of outcomes realizable by the hypothesis class, and there is an edge between two nodes representing outcomes sequences that differ on exactly one outcome. The work in \cite{predicting_01_on_randomly_drawn_points} presented an upper bound over the regret that depends upon the maximal density of any sub-graph of the one-inclusion graph. They also provided algorithms that led to this upper bound in a case where one can assign a probability for each outcome and incur the expected loss according to this probability, 
and when one has to choose an outcome. Furthermore, they showed that this maximal density is bounded by the Vapnik-Chervonenkis (VC) dimension of the class, 
\cite{vapnik2015uniform}. We note that this bound was also used to upper bound the regret of the stochastic setting.

The important work \cite{predicting_01_on_randomly_drawn_points} seems to have inspired many other works. Notably, the analysis in \cite{predicting_01_on_randomly_drawn_points} focuses on average loss, providing a PAC-bound using Markov's inequality. \cite{li2001one} have shown that their algorithm is near-optimal for the expected regret. Later, \cite{aden2023OIG_not_always_optimal} have shown that the original algorithm based on the one-inclusion graph cannot achieve optimal PAC bounds. 


Another line of work inspired by the work in \cite{predicting_01_on_randomly_drawn_points} deals with multi-class problems. In this case, \cite{Rubinstein2006Shifting} and \cite{daniely2014optimal} have shown upper and lower bounds over the $0-1$ loss based on the maximal average degree of any sub-graph of the one-inclusion hyper-graph. Building upon this work, \cite{brukhim2022characterization} defined the Daniely-Shwartz (DS) dimension and showed that it characterizes learnability for the multi-class problem.

As for the log-loss function in the individual batch learning case, for the leave-one-out regret which is the topic of this paper, \cite{Forster2002RelativeExpectedInstantaneous} has shown that for the binomial case, a simple add-$1$ learner achieves a regret of $\frac{1}{N} + o(\frac{1}{N})$ for the individual case as well as the the stochastic case. Later, \cite{Braess2002MinimaxKL} expanded this result for the multinomial case, attaining a $\frac{m-1}{N} + o(\frac{1}{N})$ bound. In a more recent work \cite{fogel2023permutation} considered the binary binomial distribution case and showed an upper bound of $\frac{1}{N} + o(\frac{1}{N})$, and showed numerically that the optimal regret is $R^*_{binary} = \frac{1}{N} - \frac{1}{2N^2}$. That work also showed a regret of order $\frac{\log(4)}{N}$ for deterministic 1-dimensional barrier threshold.

\section{Contributions and Paper Outline}
\label{section_outline}
In this paper we present and study an individual batch learning problem where the regret is constructed via a leave-one-out approach. In addition, we study the min-max optimal regret of this individual batch learning problem for $3$ different hypothesis classes:

\begin{enumerate}
    \item \emph{The multinomial (featureless) case}, where the data consists only of symbols from a finite alphabet of size $m$.
    
    For this case, we show that the min-max optimal solution is an equalizer, i.e., it induces the same regret for any outcome sequence $y^N$. We also show that the min-max optimal regret is of order $\frac{(m-1)}{N} + o(\frac{1}{N})$, and give an intuitive explanation of why the regret in the individual case is higher than the $\frac{m-1}{2N} + o(\frac{1}{N})$ in the stochastic setting.
    
    \item \emph{Deterministic hypothesis classes}, where each $x_t$ deterministically specifies $y_t$ through an underlying rule. 
    

    We analyze this case using the one-inclusion graph, \cite{predicting_01_on_randomly_drawn_points}, deriving upper bounds on the regret that depend on its properties. For the binary outcome space, we also find, for finite VC-dimension $d$, a hypothesis class whose VC-dimension is $d$ and its regret is lower bounded by $\frac{d \log(N)}{N} + o(\frac{\log(N)}{N})$.

    \item \emph{General probabilistic classes}, where $\Theta$ may represent any family of conditional distributions, where each data feature is assigned to one of two groups, and there is a fixed probability assignment given the group. The hypothesis class consists of different partitions of the data features into groups and different probability assignments.

    In this case, we show that the regret is upper-bounded by $\frac{d\log(N)}{N} + o \left( \frac{1}{N} \right)$ under relatively mild conditions. This result matches a lower bound we got in the deterministic case, thus fully characterizes the first term of the min-max regret of any hypothesis class of finite VC-dimension $d$, both for the deterministic and the stochastic hypothesis classes.

    \textbf{This result essentially validates the proposed leave-one-out regret as an eligible individual batch learning problem}, since every hypothesis class that is learnable in the classical stochastic setting, i.e., is of finite VC-dimension $d$, is also learnable for individual sequences under the leave-one-out regret.
\end{enumerate}

Finally, we discuss other individual batch-learning problems that were considered in the past, and show that the learners they induce might fail to learn a relatively simple hypothesis class, which is learnable under the leave-one-out criteria.

The structure of the paper is as follows: First, formal problem definition and general notations are formally presented in section \ref{section_definition}. Then, we follow with the analysis of the different hypothesis classes: Section \ref{section_multinomial} presents the analysis for the multinomial case, Sections \ref{section_deterministic} and \ref{section_multiclass} handle the deterministic case for binary and general-size alphabet, respectively, and Section \ref{section_probabilistic} handles the general probabilistic hypothesis classes. The comparison to previously analyzed individual batch learners is presented in \ref{pNMLs}. Finally, Section \ref{section_conclusions} concludes and proposes several possible directions for future work.

\section{Formal Problem Definition and Notations}
\label{section_definition}

Consider the case where the observation sequence $z^N = (x, y)^N$ is a deterministic, arbitrary sequence, and the regret is measured by a leave-one-out averaging:

\begin{equation}
    R_{loo} \left( z^N, \Theta, q(\cdot|\cdot) \right) = \max_{\theta \in \Theta} \frac{1}{N}\sum_{t=1}^N \log \left( \frac{p_{\theta}(y_t|x_t, z^{N \setminus t})}{q(y_t|x_t, z^{N \setminus t})} \right) 
\end{equation}
where $z^{N \setminus t} = z_1, z_2, ..., z_{t_1}, z_{t+1}, ... z_N$. Note that the probability assignment $q(\cdot|x_t, z^{N \setminus t})$ can be a function of all the examples except for the one upon which the learner is tested, for which only the data feature $x_t$ is given.  
We will be interested in the following min-max regret:

\begin{equation}
\begin{split}
    R_{loo}^*(x^N, \Theta) &= \min_{q(\cdot|\cdot)} \max_{y^N} R_{loo} \left( z^N, \Theta, q(\cdot|\cdot) \right) = \min_{q(\cdot|\cdot)} \max_{y^N} R_{loo} \left( x^N, y^N,\Theta, q(\cdot|\cdot) \right).
    \end{split}
\end{equation}
That is, we are interested in the probability assignment that attains the min-max optimal average leave-one-out regret, where the maximum is over all outcome sequences $y^N$. This min-max regret depends on both the data features $x^N$ and the hypothesis class $\Theta$, but we will usually omit them and write $R_{loo}$ or $R_{loo}^*$. 

\subsection{Notations}

Throughout the paper, we shall use the following notations:

\begin{center}
\begin{tabular}{ll}
\hline
\textbf{Symbol} & \textbf{Meaning} \\
\hline
$x_t \in \mathcal{X}$, $y_t \in \mathcal{Y}$ & Feature and outcome at time $t$ \\
$x^N, y^N$ & Sequences $(x_1,\ldots,x_N)$ and $(y_1,\ldots,y_N)$ \\
$z^N$ & Paired sequence $(x^N, y^N)$ \\
$z^{N\setminus t}$ & Sequence $z^N$ with the $t$-th pair removed \\
$\Theta$ & Hypothesis class $\{p_\theta(y\mid x): \theta\in\Theta\}$ \\
$q(y\mid x, z^{N\setminus t})$ & Learner’s predictive distribution at index $t$ \\
$R_{\mathrm{loo}}^*(x^N,\Theta)$ & Minimax leave-one-out regret (see (2)) 
\\
\hline
\end{tabular}
\end{center}

In addition, a \textbf{deterministic} hypothesis class is a hypothesis class that satisfies $\forall{\theta \in \Theta, x \in \mathcal{X}, y \in {\mathcal{Y}}: p_{\theta}(y|x) \in \{ 0, 1 \} }$. When dealing with deterministic hypothesis classes, we will say that an outcome sequence $y^N$ is \textbf{realizable} if  $\exists_{\theta \in \Theta}: p_{\theta}(y^N|x^N) = \prod_{t=1}^N p_{\theta}(y_t|x_t) = 1$.

\section{Multinomial Case}
\label{section_multinomial}
\textbf{Setting:}
Consider the multinomial case, where there are no data features, and there are $m$ possible outcomes, $y \in \{0, 1, .., m-1 \}$. The hypothesis class, in this case, will be the simplex: \[ \left\{\vec{\theta} = [\theta(0),\ldots,\theta(m-1)] \in \left[0, 1 \right] ^ m: \sum_{j=0}^{m-1}\theta(j) = 1 \right \} \] where $\theta(j)$ is the probability of the $j$-th letter according to that hypothesis. Thus, the probability of observing a sequence $y^N$ according to hypothesis $\vec{\theta}$ will be   
$p_{\vec{\theta}}(y^N) = \prod_{t=1}^N \theta(y_t)$. 

\textbf{Notation}: Denote by $\vec{v}$ the vector of empirical appearances of each letter in the observation vector $y^{N}$, such that $\vec{e} \left[  i \right]$ is the number of appearances of the letter $i$ in $y^{N}$. Naturally, the vector is of size $m$, each of its entries is a non-negative integer and they all sum to $N$. In addition, denote by $\hat{i}_j$ the vector with $1$ in the $j$-th entry and $0$ everywhere else, and by $q_{\Vec{e}}(j) = q(j|{\Vec{e}})$ the learner's probability assignment for the $j$-th letter. Note that $\vec{e}$ is a vector of appearances in the training set, so the summation over its components is $N-1$. Furthermore, when the test sample $y_t$ is some letter $j$, this letter is taken from the training set, and thus $\vec{e} = \vec{v} - \hat{i}_j$. This leads to the following simplified expression for the leave-one-out regret:

\begin{equation}
    \label{loo_multinomial_regret}
    R_{loo} \left(\vec{v}, q(\cdot|\cdot)  \right) = \sum_{j=0}^{m-1} \frac{\vec{v}(j)}{N} \log \left( \frac{\frac{\vec{v}(j)}{N}}{q(j | \vec{v} - \hat{i}_j)} \right)
\end{equation}
where $q(j|\vec{e})$ is a probability assignment for $j$ given a vector of empirical appearances $\vec{e}$. Intuitively, the $\frac{\vec{v}(j)}{N}$ term is the proportion of the letter $j$ in $v$, and thus it is also the proportion of the times $j$ is used for the test. Its appearance inside the $\log$ stems from the empirical probability being the optimal one, i.e. $\theta^* = \frac{\vec{v}}{N}$. The term $q(j | \vec{v} -\hat{i}_j)$ is the probability assigned by the learner to the letter $j$ after observing a training sequence where the number of samples of each letter is described by the vector $\vec{v} - \hat{i}_j$, which is the same as $\vec{v}$ except for the one appearance of the letter $j$ which was taken out of the training and will be used as the test.

\textbf{Main Results}: Our first result for the multinomial case follows the equalizer argument presented in \cite{fogel2023permutation} for the binary case, essentially saying that the min-max optimal solution to our problem must be an equalizer:

\begin{theorem}
\label{theorem_multinomial_equalizer}
    The min-max solution of \eqref{loo_multinomial_regret}, $R^*_{loo} = \min_{q(\cdot|\cdot)}\max_{\vec{v}} R_{loo} \left(\vec{v}, q(\cdot|\cdot)  \right)$, must be an equalizer, i.e., the regret must be equal for all possible vectors of empirical appearances $\vec{v}$.
\end{theorem}

The proof, given in the appendix, shows that given any probability assignment that does not yield the same regret across all possible training sets, one can apply a series of modifications to the probability assignment that will result in a new probability assignment with a lower maximal regret.

Our main result for this problem is a converse that states that asymptotically, one cannot achieve a min-max regret lower than $\frac{(m-1}{N}) + o(\frac{1}{N})$.

\begin{theorem}
\label{theorem_multinomial_regret}
    The min-max optimal leave-one-out regret of the multinomial problem satisfies:
    
    \begin{equation*}
        R_{loo}^* \geq \frac{(m-1)}{N} + o \left( \frac{1}{N} \right).
    \end{equation*}
\end{theorem}

To prove this Theorem, we consider sequences that consist almost entirely of a specific letter, for example, nearly all $0$'s sequence. If the regret is smaller than the lower bound, say $R_{loo}^* = \frac{(m-1)(1-\epsilon_0)}{N} + o(\frac{1}{N})$ it implies that it is smaller for the all $0$'s sequence, and so we get that the assigned $q$ for the symbol $0$ given the rest $N-1$ zeros must be larger than some value. This immediately implies an upper bound on the probabilities assigned to the other symbol values, given the $N-1$ zeros. Now, for a sequence with a single non-zero letter, we can utilize this upper bound and the fact that the regret for this single non-zero sequence is also $\frac{(m-1)(1-\epsilon_0)}{N} + o(\frac{1}{N})$ to get another lower bound on the average probability assigned to zero, given a training set with $N-2$ zeros and a single non-zero letter, by $\frac{m(1-\epsilon_1)}{N} + o(\frac{1}{N})$. We can continue recursively, adding another non-zero symbol, and lower-bound the average probability assigned to zero after seeing $k$ non-zero letters by $1 - \frac{(m+k-1)(1-\epsilon_k)}{N}$. Furthermore, we show that at some point $\epsilon_k \geq 1$, which implies that some assigned probabilities must be zero. For log-loss, this leads to an infinite regret for the relevant sequences, which leads to a contradiction.


The full proof naturally consists of an induction over the number of letters other than $0$ in the training set and is given in the appendix. It also involves solving a relatively simple constrained optimization problem to deal with the learners' various possibilities to assign probabilities for the non-frequent letters. In addition, it utilizes Jensen's inequality to bound the average probabilities of the frequent letter given a bound over the average log-loss, which is the logarithm of the geometrical average.

This lower bound also has a matching upper-bound over the regret up to $o(\frac{1}{N})$ terms, which is attained by the 'add-$1$' estimator - see \cite{Braess2002MinimaxKL}. Thus we can conclude that the min-max optimal regret for the multinomial case is indeed $\frac{(m-1)}{N} + o(\frac{1}{N})$.

Our last theoretical result for this case analyzes the 'add-$1$' estimator and show it attains a constant regret up to $o(\frac{1}{N^2})$ terms:

\begin{theorem}
\label{theorem_add_1_regret}
 The regret of the probability assignment $q_{\vec{e}}(j) = \hat{\theta}_j + \frac{1}{N-1} \left( 1 -m\hat{\theta}_j \right)$ is:
 \begin{align*}
    \begin{split}
    &R_{loo} \left(\vec{v}, q_{\vec{e}}(j) 
    = \hat{\theta}_j + \frac{1}{N-1} \left( 1 -m\hat{\theta}_j \right)\right) =  \frac{m-1}{N} - \frac{(m-1)^2}{2(N-1)^2} + o(\frac{1}{N^2}).
    \end{split}
 \end{align*}
 where $\hat{\theta}_j$ is the rate of appearance of the $j$-th letter in the training set, $\frac{\vec{e}(j)}{N-1}$.
\end{theorem}

The full proof is given in the appendix and is quite simple, building upon the first two terms of Taylor's expansion of $\log(1-x)$. Note that using Taylor's expansion for $\frac{1}{1 + x}$ one can show that this estimator is equivalent to the add-$1$ estimator up to $o(\frac{1}{N^2})$. It is interesting to note that this probability assignment not only achieves the $\frac{m-1}{N}$ lower bound but also attains an equalizer for the second term, which might hint that this is the optimal solution up to $o(\frac{1}{N})$ terms.

\textbf{Comparison to the Stochastic Case}: Note that for the stochastic case, the regret is $\frac{(m-1)}{2N} + o(\frac{1}{N})$, and thus we have a factor $2$ difference in the leading term between the stochastic and the individual settings.

The intuitive reason for the difference between the two settings is also of interest: When $\forall_{j}\;\; N \cdot \theta(j) \gg 1$, then we can apply the central limit theorem and get for the stochastic setting that the difference between the empirical probability in the training set, upon which the learner bases its prediction, deviates from the true probability by approximately $\frac{\theta(j) (1-\theta(j))}{\sqrt{N}}$. On the other hand, for the individual learning, the difference between the empirical probability the learner observes and the 'true' empirical probability is bounded by $\frac{1}{N}$: Indeed, when asked about a probability for a letter $j$, the learner has observed $\vec{v}(j) -1$ appearances of $j$ out of $N-1$, while the probability to which the learner will be compared, is, of course, $\frac{\vec{v}(j)}{N}$ - and the difference is $\frac{N - \vec{v}(j)}{N}\frac{1}{N-1}$. So, how can a closer-to-reality estimate lead to a worse regret?

The answer to this question is that, for the individual case, the empirical distribution of $j$ observed by the learner when the test outcome is $j$ is always lower than the 'true distribution' and thus does not sum up to one. On the other hand, the empirical distribution in the stochastic case is always valid. To see why this leads to the difference, consider a learner who uses the empirical probability. In the individual case, if we denote by $\theta(j) = \frac{\vec{v}(j)}{N}$ the 'true distribution,' we will get that the learner assigns to the $j$-th letter $\hat{\theta}(j) = \theta(j) - \frac{1 - \theta(j)}{N-1}$. Assuming $\forall_j \theta_j >> \frac{1}{N}$, we get:

\begin{align*}
    &R = \sum_{j} \theta(j) \log \left( \frac{\theta(j)}{\hat{\theta}(j)} \right) 
    = \sum_j -\log \left( 1 - \frac{1 - \theta(j)}{(N-1)\theta(j)} \right) 
    = \sum_j(\frac{1 - \theta(j)}{N-1}) + o(\frac{1}{N}) = \frac{m-1}{N} + o(\frac{1}{N}).
\end{align*}

Note that the leading term here comes from the first term of $-\log(1-x) = x + \frac{x^2}{2} + o(x^2)$. On the other hand, for the stochastic case, if we denote by $n_j$ the number of $j$'s in the training set, and $ \delta_j = \frac{n_j}{N} - \theta_j$ the difference between empirical and true probabilities, we get $\sum_j \delta_j = \frac{N}{N} - \sum_j \theta_j = 0$, and thus:

\begin{align*}
    &R 
    = \sum_j \theta_j \log \left( \frac{\theta(j)}{\frac{n_j}{N}} \right) 
    = -\sum_j \theta_j \log \left( 1 + \frac{\delta_j}{\theta_j} \right)
    = \sum_j -\delta_j + \frac{\delta_j^2}{2 \theta(j)} + o(\delta_j^2) 
    = \sum_j \frac{\delta_j^2}{2 \theta(j)} + o(\frac{1}{N}).
\end{align*}

To conclude, even though the observed empirical distribution is closer to the ``true distribution'' in the individual case, the bias in the empirical distributions in this case forces a larger regret.

\section{Deterministic Hypothesis Classes}
\label{section_deterministic}
\textbf{Setting:}
Consider now a case where there are data features $x$, the outcome space is binary, e.g. $y \in \{0, 1\}$, and the hypothesis classes are deterministic functions, i.e., $p_\theta(y|x) \in \{0, 1 \}$. In this case, we say that some sequence $z^N = (x, y)_{t=1}^N$ is realizable if there is some $\theta \in \Theta$ such that $\prod_{t=1}^N p_{\theta}(y_t|x_t) = 1$.

\textbf{The one inclusion graph}: We can formulate the leave-one-out problem with log-loss using the one inclusion graph \cite{predicting_01_on_randomly_drawn_points} as follows: Given a sequence $x^N$, construct a graph where each node describes a realizable sequence, and there is an edge between each pair of nodes that differ only on a single outcome $y_t$. Each edge represents a probability that the learner has to assign $q(\cdot|z^{N \setminus t}, x_t)$, and the regret associated with a specific realizable $y^N$ is the summation over all connected edges of minus the logarithm of the probabilities associated with the $y_t$ of that node. Note that for the $0-1$ loss with randomized predictions, one would have had to replace the minus-logarithm of the probability with $1$ minus the probability.

\begin{figure}
    \centering
    \includegraphics[width=1\linewidth]{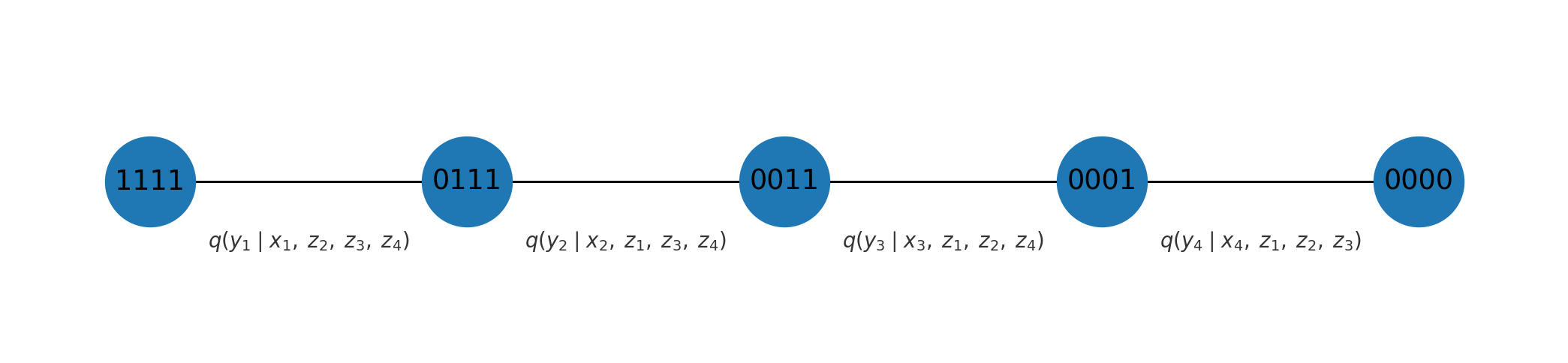}
    \caption{One inclusion graph for the deterministic hypothesis class of 1-dimensional barrier threshold, $n=4$, $x_0 < x_1 < x_2 < x_3  < x_4$.}
    \label{1-inclusion graph 1-dimensional barrier threshold}
\end{figure}

Figure \ref{1-inclusion graph 1-dimensional barrier threshold} demonstrates how our problem can be cast into assigning probabilities for the one inclusion graph. Here, we use the 1-dimension barrier threshold where \[
p_{\theta}(y \mid x)
=
\begin{cases}
1, & \text{if } (y=1 \text{ and } x>\theta)\ \text{or}\ (y=0 \text{ and } x\le\theta),\\[4pt]
0, & \text{otherwise.}
\end{cases}
\] 
and assume $n=4$ and $x_1 < x_2 < x_3 < x_4$. Each node represents one of the $5$ realizable sequences, while each edge connects two nodes that represent the same sequence up to exactly one outcome $y_t$, and thus it is associated with the probability assignment for $q(\cdot | x_t, z^{N \setminus t})$.

\textbf{Results:}
Our first observation is that if the one-inclusion graph is connected, any min-max optimal probability must also be an equalizer:

\begin{theorem}
\label{theorem_equalizer_deterministic_vc_dimension}
    Assuming that the one-inclusion graph is connected, the min-max optimal solution for the deterministic hypothesis class must be equal for all realizable sequences $y^N$.
\end{theorem}

The proof is similar to the multinomial case and appears in the appendix. The interesting part here is that this property might not hold for all loss functions. Consider the $0-1$ loss function, which has been analyzed in \cite{predicting_01_on_randomly_drawn_points}. In the $0-1$ case, the summation of the regret over all sequences is simply the number of edges divided by $N$, regardless of the probability assignment. To see this, note that the probabilities in each edge sum to one, and since the regret of a node is a simple summation of terms of the form $1-q$, we get that the sum of the regrets over all sequences is just $\frac{E}{V \cdot N}$, where  $E$ and $V$ are the numbers of edges and nodes of the one inclusion graph.

Thus, if the optimal probability assignment would have led to an equal regret for every realizable $y^N$, we would have gotten a regret that depends on the density of the one-inclusion graph and not on the maximal density of any sub-graph. Alas, to show that this is not always achievable, we observe that the regret of a specific node is always lower than the number of edges connected to it. Thus, if there is a node whose number of edges is lower than the density of the graph, then its regret will be lower. Since the summation over the regrets does not depend on the probability assignment, there must be a node whose regret is higher, and thus, the min-max regret is higher than the density. Another way to look at it is that we cannot fully utilize all nodes to divide the regret equally.

The reason the solution must be an equalizer if the one inclusion graph is connected for the log-loss is that the log-loss is unbounded. Thus, given some probability assignment, if at least one node has a regret which is strictly lower than the maximal regret, one can always find an edge between a node with the maximal regret and a node with strictly lower regret. One can then alter the probability that affects only those two nodes to equalize between them. One can repeat this step if the regret is still not equal between all nodes.

Next, we shall consider a bound over the regret, which depends on the minimal degree of the sub-graphs of the one-inclusion graph:

\begin{theorem}
    \label{theorem_max_min_degree_regret_bound}
    Assume that for some deterministic hypothesis class $\Theta$ and data features sequence $x^N$, the minimal degree of any sub-graph of the one-inclusion graph is some $k$. In this case: 
    \begin{align*}
        R^*_{loo} \leq \frac{k \log(N)}{N} + o\left( \frac{\log(N)}{N} \right)
    \end{align*}
\end{theorem}

The full proof, given in the appendix, is a constructive proof, where a probability assignment is recursively defined: At each step, we define the outer layer as the group of all nodes whose degree is at most $k$. We assign a probability of $\frac{1}{2}$ to edges connecting two nodes in the outer layer and a probability of $\frac{1}{N}$ for edges between the outer layer and nodes outside of it.  

To bound the regret of a specific node, we divide the edges into three groups. Edges that were dealt with when the node was not in the outer layer; edges that were dealt with when the node was in the outer layer and connected to a node in the same layer; and edges that were dealt with when it was in the outer layer and connected the node to a node in an inner layer. We show that the latter group is dominant in the regret analysis, which yields the $\frac{k \log(N)}{N}$ term.

Arguably, one of the most fundamental technical cornerstones that led to the bound over the leave-one-out regret for the $0-1$ loss in \cite{predicting_01_on_randomly_drawn_points} is that the VC dimension of the hypothesis class upper bounds the maximal density of any sub-graph of the one-inclusion graph. One can easily combine this result and Theorem \ref{theorem_max_min_degree_regret_bound} to come up with the following bound:

\begin{corollary}
\label{corollary_vc_dimension}
    For every deterministic hypothesis class of finite VC-dimension $d$, and every sequence $x^N$, the following holds:
    \begin{equation*}
        R^*_{loo} \leq \frac{2d \log(N)}{N} + o \left( \frac{\log(N)}{N} \right)
    \end{equation*}
\end{corollary}


On first look, the upper bound we got over the min-max regret might seem a bit disappointing because for the $0-1$ loss \cite{predicting_01_on_randomly_drawn_points} got a bound of $\frac{d}{N}$, and we also got a bound that scales like $\frac{(m-1)}{N}$ with log-loss for the multinomial case. Thus, one may wonder if the bound can be improved to the form $\frac{c}{N}$. Our following result provides sufficient conditions under which the leading term is of order $\frac{c}{N}$:

\begin{theorem}
    \label{theorem_max_degree_regret_bound}
    If the maximal degree of the one inclusion graph is bounded by $k$, then:
    \begin{align*}
        R_{loo}^* \leq  \frac{k}{N}
    \end{align*}
\end{theorem}

The proof of this lower bound is straightforward - assign a probability of $\frac{1}{2}$ in each of the edges, and the bound follows immediately.

Here, it is essential to note that there are hypothesis classes of finite VC-dimension $d$ where the maximal degree of the one-inclusion graph is $N$. One such example, which we call ``$d$-unique-values'', is the following family of hypothesis classes: The data features are $x \in [ 0, 1 ]$, and every hypothesis $\theta \in [0, 1]^d$ assigns $y=0$ for all values of $x$ except $d$ specific values $\theta_1, \theta_2, ... \theta_d$, for which $1$ is assigned. This example has appeared in \cite{predicting_01_on_randomly_drawn_points} for $d=1$, a case for which the node representing $y^N = 0^N$ has $N$ edges. An example of the one-inclusion graph for $d=2, N=5$ is presented in  Figure \ref{1-inclusion graph for 2-unique values, N=4}.

\begin{figure}
    \centering
    \includegraphics[width=1\linewidth]{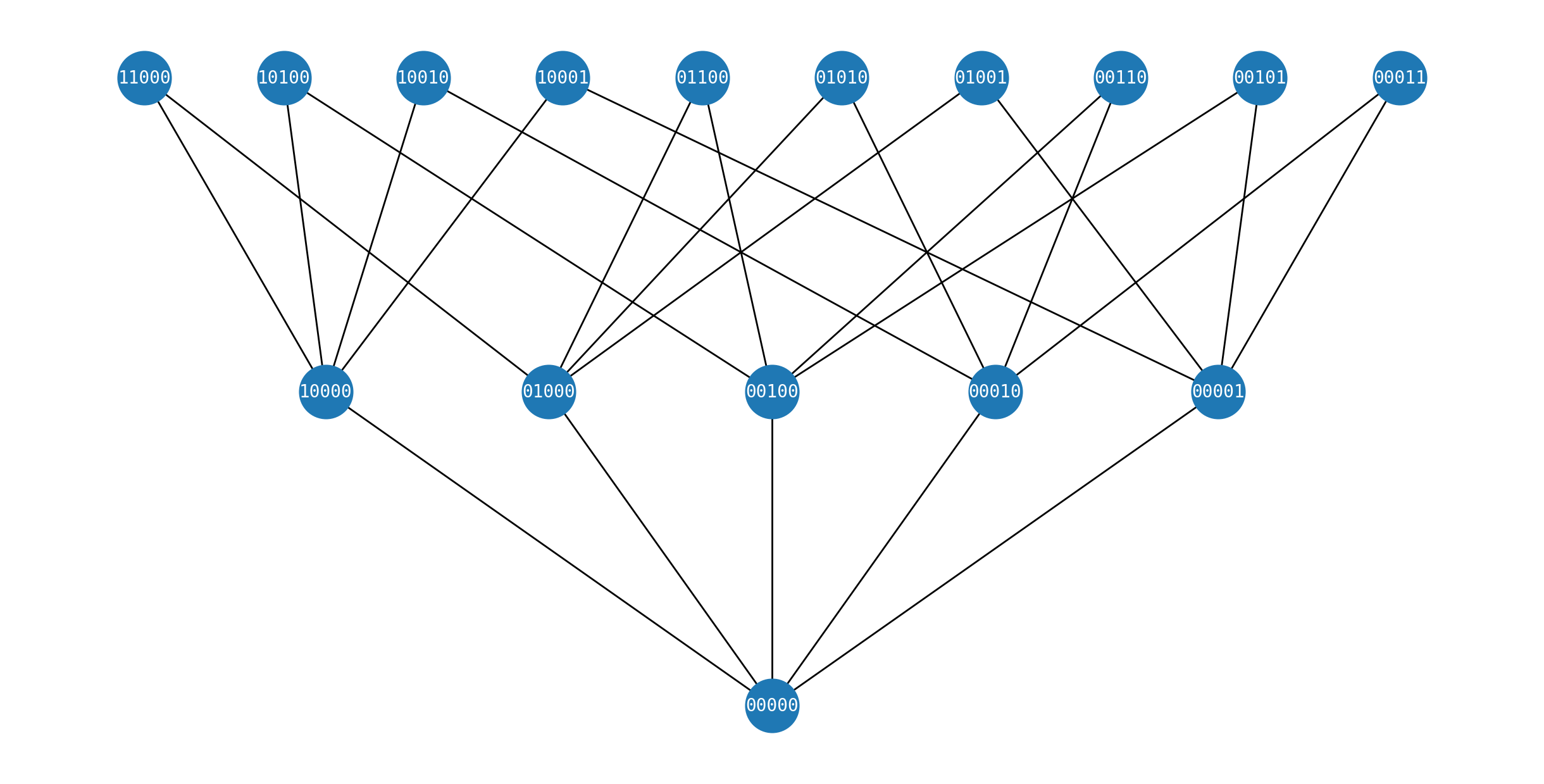}
    \caption{One inclusion graph for $2$-unique values, $N=5$. The bottom node represents an all-$0$ sequence, the middle layer contains nodes representing sequences with a single $1$, while the sequences in the upper layer contain $2$ $1$-s.}
    \label{1-inclusion graph for 2-unique values, N=4}
\end{figure}

It turns out that these specific hypothesis classes can also be utilized to get the following lower bound on the regret:

\begin{theorem}
    \label{theorem_d-unique_values_regret}
    For every finite VC-dimension $d$, there is a hypothesis class for which the min-max optimal regret is $ R_{loo}^* \geq \frac{d \log(N)}{N}  + o \left( \frac{\log(N)}{N} \right)$.
\end{theorem}


To conclude, note that the minimal degree of every sub-graph of the ``$d$-unique-values'' is indeed $d$; This can be easily verified by observing that the degree for every node with $d$ ones is $d$ and that if we remove those nodes, we get the one-inclusion graph for ``$d-1$-unique-values''. Thus, the technique described in \ref{theorem_max_min_degree_regret_bound} indeed attains the optimal regret of $\frac{d \log(N)}{N}$ up to $o \left(\frac{\log(N)}{N} \right)$ terms. 

\section{Multi-class}
\label{section_multiclass}
\textbf{Setting}: Consider now a generalization of the deterministic hypothesis classes to a multi-class scenario, where there are $m$ possible outcomes $y \in \{ 0, 1, ..., m-1\}$. Again, we will assume that the hypothesis class is deterministic. 

\textbf{Results}: In \cite{daniely2014optimal}, it was shown that if the average degree of any sub-graph of the one-inclusion hyper-graph is bounded by some $\mu$, then the $0-1$ loss is between $\frac{\mu}{2N}$ and $\frac{\mu}{N}$. Here, we will show that such a bound over the maximal average degree also leads to a vanishing leave-one-out regret with the logarithmic loss function.  

\begin{theorem}
\label{theorem_multiclass}
    If the maximal average degree of any sub-graph of the one-inclusion hyper-graph of the hypothesis class $\Theta$ is upper-bounded by some $\mu$, then:
    \begin{align*}
        R^*_{loo} \leq \frac{\mu \log(N)}{N} + o \left( \frac{\log(N)}{N} \right).
    \end{align*}
\end{theorem}
The full proof of this theorem also appears in the appendix. Similar to the proof for the binary case, we utilize the bound over the average degree of any sub-graph of the one-inclusion hyper-graph. Here also, we define the outer layer as the group of all nodes whose number of edges is not greater than this bound, which is now $\mu$. The main obstacle here is that an edge contains several nodes, and thus, we might have to assign the probabilities for it over several layers. Thus, we will do so as follows: For each node in the outer layer, consider each connected edge - if there are other nodes from inner layers in the edge, assign a probability of $\frac{1}{(m-1) \cdot N}$; otherwise if the only remaining nodes in the edge are from the outer layer, assign all the remaining probability to all remaining nodes equally. A similar analysis to the binary case reveals that the dominant term in the regret of any node comes from the edges that included nodes from either the same layer or an inner layers. This bound is of order $\frac{\mu \log(N)}{N} + o \left( \frac{\log(N)}{N} \right)$. 

\section{Probabilistic Hypothesis Classes}
\label{section_probabilistic}
\textbf{Setting}: Consider now a more complicated case where the hypothesis class consists of conditional probabilities: $\forall_{\theta, x, y } p{_\theta}(y|x) \in \left[ 0, 1\right]$ and $\forall_{\theta, x} \sum_{y \in Y} p(y|x) = 1$. More specifically, we will consider hypothesis classes where each $x$ is assigned into one of two groups $j \in \{0, 1 \}$, and each group assigns a different probability for $y$. Thus, if hypothesis $\theta$ assigns $x$ into group $j=g(\theta, x)$, then $p_\theta(y|x) = p_{g(\theta, x)}(y)$. Such hypothesis classes have been studied for online learning in \cite{fogel2017problem} and \cite{bhatt2021sequential}. 

\textbf{Results}: In this case, under some rather mild conditions, one can derive the following upper bound over the regret:

\begin{theorem}
\label{theorem_probabilistic_vc_dimension}
   If $x^N$ contains $N$ different values, the leave-one-out regret for a hypothesis class $\Theta$ that partitions the samples into two groups and assigns a probability $p_j(y), j \in \{0, 1\}$ given each group, is upper bounded by:
    \begin{equation*}
        R_{loo}^* \leq \frac{ d\log(N)}{N} + o(\frac{1}{N})
    \end{equation*}
    where $d$ is the VC-dimension of the partitioning of the samples into the two groups.
\end{theorem}

The full proof appears in the appendix. The proof utilizes Sauer–Shelah lemma, due to \cite{sauer1972density} and \cite{shelah1972combinatorial}, to bound the number of possible partitions of $N$ examples into groups by $|g(x^N)| \leq (e \cdot N)^d$, and thus convert this problem to that of predicting with $K \leq (e \cdot N)^d$ experts' advice. We then define a learner based on a mixture model for the experts' prediction in a way that allows us to utilize results by \cite{kivinen1999averaging} that bound the regret between this mixture learner and the best expert in hindsight by $\frac{\log(K)}{N}$, where $K$ is the number of experts.

\textbf{Discussion}: Interestingly, we got for the more challenging case of non-deterministic hypothesis classes, a bound of order $\frac{d \log(N)}{N} + o\left( \frac{\log(N)}{N} \right)$. Naturally, one can also use this method for binary deterministic hypothesis classes. This, alongside the lower bound over the min-max optimal regret for the ``$d$-unique-values'' hypothesis classes, actually provides matching lower and upper bounds of $\frac{d \log(N)}{N} + o \left( \frac{\log(N)}{N} \right)$ for both the stochastic and deterministic hypothesis class of finite VC-dimension $d$ for the binary outcome case.

\section{Comparison To Other Individual Batch Learning Problems}
\label{pNMLs}

As we have mentioned in the introduction, two other possible formulations for the individual batch-learning problem have been considered in past works. In this section, we will briefly introduce them and show that there is a simple, learnable hypothesis class where they fail to learn, i.e., their resulting regret is constant regardless of $N$.

In \cite{fogel2019individual}, we have proposed to fix the training set in advance and test both the learner and the reference on the single test set. In this case, a reasonable reference may also know the test set in advance, but it will not which of the examples is in the training and which is in the test. Thus, we get the following min-max problem:

\begin{align*}
    \label{min_max_pNML1}
    &R^*(x_N,z^{N-1}) = \min_q \max_y R(q,x_N,y_N,z^{N-1}) \log\left(\frac{p_{\hat{\theta}(z^N)}(y_N|x_N)}{q(y_N|x_N;z^{N-1})}\right)
\end{align*}

The min-max probability assignment is:

\begin{align}   
q_{\mbox{\tiny{pNML}}}(y_N|x_N; z^{N-1}) = \frac{p_{\hat{\theta}(z^N)}(y_N|x_N)}{\sum_{\Tilde{y}_N}p_{\hat{\theta}(\Tilde{z^N})}(\Tilde{y}_N|x_N)}
\end{align}
where $\theta(z^N) = \arg \max_{\theta} p_{\theta}(y^N|x^N).$

Another possible formulation was discussed in \cite{fogel2022another}, where again, one assumed that the training set is fixed in advance and the maximization is other than the test outcome. The difference was that we measured both the learner and the reference over all the examples:

\begin{align}
    \label{Regret_perm_2}
    R_{perm} = \min_{q(\cdot| x^N, y^{N-1})} \max_{\theta, y_N}
    \log(\frac{p_{\theta}(y^N|x^N)}{q(y_N|x^N, y^{N-1})})
\end{align}

If we assume that the hypothesis class contains only probability assignments of the form $p_{\theta}(y^N|x^N) = \prod_{t=1}^Np_{\theta}(y_t|x_t)$, and that there are no repetitions among the data-features $x^N$, then one can show as in \cite{fogel2022another} that this leads to:

\begin{align}   
q_{\mbox{\tiny{pNML-2}}}(y_N|x_N; z^{N-1}) = \frac{p_{\hat{\theta}(z^N)}(y^N|x^N)}{\sum_{\Tilde{y}_N}p_{\hat{\theta}(\Tilde{z^N})}(\Tilde{y}^N|x^N)}.
\end{align}

The two derived solutions have been previously studied in \cite{roos2008sequentially}, \cite{Grunwald2007}, and \cite{roos2008bayesian} as approximations for the NML distribution with two main advantages: First, these approximations can be computed more efficiently. Second, there are cases where the denominator of the NML diverges, and thus, the NML is undefined, whereas the approximations can be calculated. We denoted it in previous works by pNML and pNML-2 to emphasize the motivation of a universal batch learner.

On top of having a closed-form expression for the general hypothesis class case and also more concrete expression for different hypothesis classes - see \cite{bibas2019new} and \cite{fogel2022another}, these two solutions also have an algorithmic interpretation that can be used for general learning algorithms: Given the test data feature, we take every possible outcome, add it to the training set and run the learning algorithm. For the pNML, we take the derived probability from the test sample and normalize it. On the other hand, for the pNML-2, we take the derived probability over all the samples and normalize. Naturally, for deep neural networks, where training the network from scratch for all possible outcomes is prohibitive, we can take a few iterations for every possible label - see \cite{bibas2019deep} for example. This method has also shown promising results in \cite{fu2021offline} and \cite{li2021mural}.

Despite having promising practical implementations, it turns out that there are simple, learnable hypothesis classes where both the pNML and pNML2 fail to learn, i.e. there is a sequence for which the regret is constant and does not converge to $0$ as $N \to \infty$:

\begin{theorem}
\label{theorem_pNMLs_fail}
  For the "1-unique-value" hypothesis class, where $y=1$ only if $x=\theta$ and $0$ otherwise, the regret of both the pNML and pNML-2 is $\log(2)$ for $y^N = 0^N$ when $x^N$ contains $N$ different values. 
\end{theorem}

The proof is rather straight-forward and given in the appendix.

\section{Conclusions and Future Work}
\label{section_conclusions}
In this work, we analyze a variant of batch learning with log-loss, where the outcome sequence is deterministic and the regret is measured using a leave-one-out approach.

For the case of the multinomial distribution, we found that the regret is lower bounded by $\frac{(m-1)}{N} + o\left(\frac{1}{N}\right)$. This lower bound matches a known upper bound of $\frac{(m-1)}{N} + o\left(\frac{1}{N}\right)$, thus fully characterizing the first order of the regret. We also discussed why in this case the regret is larger than the $\frac{m-1}{2N} + o\left(\frac{1}{N}\right)$ optimal regret in the stochastic batch learning problem, even though the empirical distribution of the training set is much closer to the 'true' distribution in the individual setting than in the stochastic setting.

For general hypothesis classes in which each hypothesis assigns the data feature to one of two groups, and then assigns a probability for $y$ given the group, we showed that the regret is upper bounded by $\frac{d \log(N)}{N} + o \left( \frac{\log(N)}{N} \right)$, where $d$ is a the VC-dimension of the set of partitions of the data features to groups. We further show that for every $d$ there is a hypothesis class of VC-dimension $d$ whose regret is $\frac{d \log(N)}{N} + o \left( \frac{\log(N)}{N} \right)$. For deterministic hypothesis classes, we provide additional upper bounds which depend upon the one-inclusion graph, and generalize our results for the multi-class setting.

This work leaves several open questions as future work directions. First, for the multinomial case, it would be interesting to see if the $ -\frac{(m-1)^2}{2(N-1)^2}$ is the min-max optimal second term. This might be supported by the fact that this is an equalizer of the second term of the regret, and it is achievable by using the probability assignment $q_j = \hat{\theta}_j + \frac{1 - m \hat{\theta_j}}{N-1}$.

Another open question is whether one can find better bounds for either the deterministic or the stochastic hypothesis class cases, perhaps conditioned on some properties of the partitioning of the data features into groups. It would be especially interesting if one could come up with some conditions over a stochastic hypothesis class for which a bound of order $\frac{1}{N}$ over the regret will be achieved. In addition, it would be interesting to attain the same result without assuming the data features are all different.

Another different line of work that might be of interest is to generalize the batch setting to a case where instead of a single test sample there are several, say $k>1$ examples, used as a test set.






%

\begin{appendices}
\section{Proofs}


\begin{IEEEproof}[Proof of Theorem \ref{theorem_multinomial_equalizer}]
    This proof is a generalization of the equivalent Theorem for the binary case that appears in \cite{fogel2023permutation}.

\textbf{Notations}:    For the readers' convenience, we shall repeat the notations from section \ref{section_multinomial} that we shall use for this proof: Denote by $\vec{v}$ the vector of size $m$ of empirical appearances of each letter in the observation vector $y^{N}$, such that $\vec{e} \left[  i \right]$ is the number of appearances of the letter $i$ in $y^{N}$. In addition, denote by $\hat{i}_j$ the vector with $1$ in the $j$-th entry and $0$ everywhere else, and by $q_{\Vec{e}}(j) = q(j|{\Vec{e}})$ the learner's probability assignment for the $j$-th letter. Note that $\vec{e}$ is a vector of appearances in the training set, so the summation over its components is $N-1$. Furthermore, when the test sample $y_t$ is some letter $j$, this letter is taken from the training set, and thus $\vec{e} = \vec{v} - \hat{i}_j$. This leads to the following simplified expression for the leave-one-out regret:

\begin{equation}
    R_{loo} \left(\vec{v}, q(\cdot|\cdot)  \right) = \sum_{j=0}^{m-1} \frac{\vec{v}(j)}{N} \log \left( \frac{\frac{\vec{v}(j)}{N}}{q(j | \vec{v} - \hat{i}_j)} \right)
\end{equation}
Where $q(j|\vec{e})$ is a probability assignment for $j$ given a vector of empirical appearances $\vec{e}$; We will sometimes use the notation $ q_{\vec{e}}(\cdot) = q(\cdot | \vec{e})$.

In addition, we will use the following notation: we say that two vectors of empirical number of appearances $\vec{v}, \vec{v}'$ are $k,l$-connected if the number of samples in each letter is equal in both, except for the two coordinated $k, l$ where the difference is one, or more explicitly: $\vec{v} - \vec{v}' = \hat{i}_k - \hat{i}_l$. Note that if we draw a graph where the possible vectors of an empirical number of appearances are the nodes and there are edges between every two nodes that are connected by some $j, l$, we get a connected graph. 

This notation is demonstrated in Figure \ref{Empirical_Number_of_Appearances_Graph} for $m=3, N=5$. The $21$ nodes represent the different vectors of empirical appearances, and an edge connects each pair of nodes if they represent sequences that can differ on a single outcome. The edges thus represent the probabilities assigned to that different outcome given the same empirical appearances in the training set.

\begin{figure}
    \centering
    \includegraphics[width=1\linewidth]{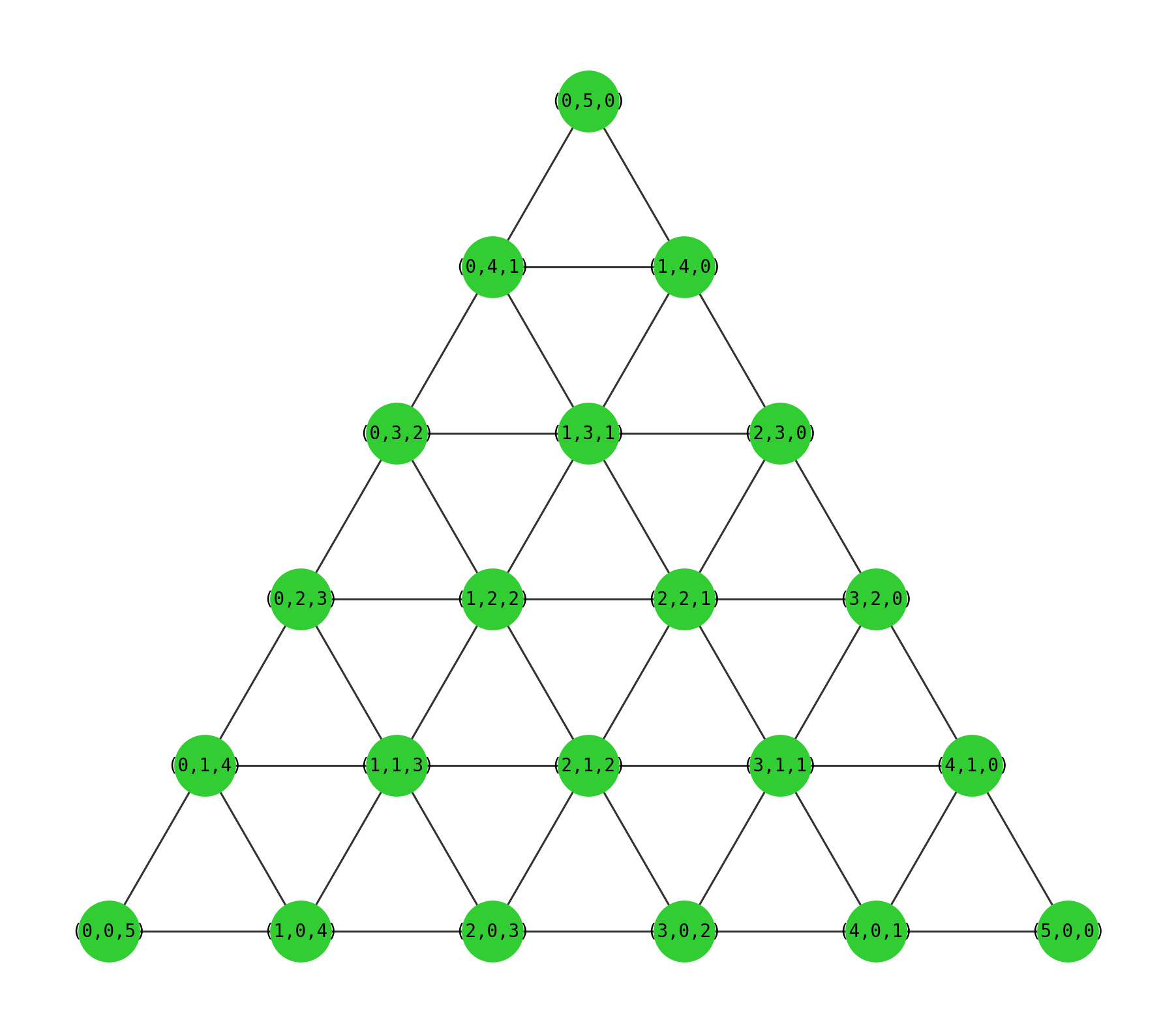}
    \caption{Empirical Number of Appearances Graph, $m=3, N=5$. Note that the three edges connecting the three upper nodes represent the probability assignments for $q \left( \cdot| \vec{e} = (0, 4, 0) \right)$. If we increase the probability assigned to $1$ at the expanse of the probability assigned to $2$ while keeping the probability assigned to $3$ constant, it will decrease the regret associated with $\vec{v} = (1, 4, 0)$, increase the regret associated with $\vec{v} = (1, 5, 0)$, and will not influence the regret associated with $\vec{v} = (0, 4, 1)$.}
    \label{Empirical_Number_of_Appearances_Graph}
\end{figure}

\textbf{General Idea:} The outline of the proof is as follows: We assume some probability assignment $q(\cdot | \cdot)$ is the min-max optimal learner and that the regret it induces is not equal for every possible training set represented by $\vec{v}$. It is sufficient to consider only probability assignments $q(\cdot|\cdot)$ that are greater than zero for every training set $\vec{e}$ and possible letter $j$, because otherwise, the maximal regret will be infinite.  If the maximal regret is achieved at a single $\vec{v}$, we show that a modified probability assignment can yield a lower maximal regret. If the maximal regret is achieved at several values of $\vec{v}$, say some $k > 1$, we show that we Then, we show that we can modify this probability assignment in a way that will lower the number of values of $\vec{v}$ for which the regret is maximal. Combining these two observations, we conclude that the maximal regret of any probability assignment that does not yield the same regret for all $\vec{v}$ can be improved.

\textbf{Single Maxima Case}: Consider probability assignments $\hat{q}_{\vec{e}}(\cdot)$ for which the regret $R(\vec{v}, \hat{q}_{\vec{e}}(\cdot))$ attains its maximal value at a single $\vec{v}$. Now, take some $v'$ that such that $v$ and $v'$ are $k, l$-connected. Denote by $\vec{e} = \vec{v} - \hat{i}_k = \vec{v}' - \hat{i}_l$, and by $\beta = \sum_{j \neq k, l} q(j | \vec{e})$ the sum of probabilities assigned to all the other symbols given the same training set $\vec{e}$. Now, we shall consider the set of all probability assignments $q_{\alpha}$ that are identical to $\hat{q}$ for every training set other than $\vec{e}$, and for the training set $\vec{e}$ assign the same probability for letters other than $k$ and $l$; $\alpha$ and  $\beta - \alpha$ are the probability assigned to $k$ and $l$ given $\vec{e}$, respectively. Now, the regret of $q_{\alpha}$ associated with$\vec{v}$ decreases as $\alpha$ increases, while the regret of $q_{\alpha}$ associated with $\vec{v}'$ increases as $\alpha$ increase up to $\infty$ for $\alpha = \beta$ - both in a continuous dependency. Also note that for the $\alpha \in (0, \beta)$ corresponding to the original probability assignment $\hat{q}$ the regret associated with $\vec{v}$ is strictly larger than that associated with $\vec{v}'$. Thus, by the intermediate value theorem, we are guaranteed that we can find $\alpha$ such that the two regrets will be equal, and we get a new probability assignment whose maximal regret is lower than that of $\hat{q}$,.

\textbf{Multiple Maxima Case}: Next, we show that if for some $\hat{q}_{\vec{e}}(\cdot)$ the maximal regret is attained at some of the vectors $\vec{v}$ but not at all of them, we can adjust $\hat{q}$ to make the maximal regret to be the same, but it will be attained at a smaller number of vectors $\vec{v}$. Indeed, in this case, since the graph of the empirical appearances vector is connected, there must be some vector $\vec{v}$ for which the maximal regret is attained, which is connected to a vector with a strictly lower regret. Utilizing the argument from the previous paragraph, we can adjust the probability assignment so that the regrets of these two connected vectors will be equal, without affecting the regret of any other node, thus lowering the number of vectors for which the maximal regret is attained by 1. 

\textbf{Every Non-Equalizing Probability Assignment Can Be Improved}: We can continue repeating this until there is only one vector where the maximal regret is attained. This leads us back to the scenario we dealt with in the previous paragraph, where we showed that we can reduce the maximal regret. Thus, we conclude that any $\hat{q}_{\vec{e}}(\cdot)$ for which the regret is not equal for all $\vec{v}$ is not the min-max optimal solution, because we can find another probability assignment whose maximal regret is strictly smaller. 
    
\end{IEEEproof}

\begin{IEEEproof}[Proof of Theorem \ref{theorem_multinomial_regret}]
    
\textbf{Notations}: Before proving the Theorem, it will be beneficial to clarify some notations. First, we will define an observation vector as a vector $\vec{e}$ whose $j$-th entry represents the number of times the letter $j$ appeared in the relevant training. Thus, each entry of $\vec{e}$ is a non-negative integer, and $\sum_{j=0}^{m-1}\vec{e}(j) = N-1$. Denote by $E_{k}$ the set of all observations vectors for which $\vec{e}(0) = N - k - 1$. Since the learner sees $N-1$ observations, for every $\vec{e}\in E_k, \;\sum_{j=1}^{m-1} \vec{e}(j) = k$. Also, denote by $q_{\vec{e}}(i)$ the learner's probability assignment for the letter $i$ after observing a training set characterized by the observation vector $\vec{e}$.

Similarly, we will denote by $R_{\vec{v}}$ the regret associated with a sequence where the $i$-th letter appears $\vec{v}(i)$ times; Note that every entry of $\vec{v}$ is also a non-negative integer, and $\sum_{j=0}^{m-1}\vec{v}(j) = N$. The group $V_k$ will consists of all such vectors $\vec{v}$ with $\vec{v}(0) = N - k$.

If we denote by $\hat{i}_j$ the $m$-dimensional vector with 1 in the $j$-th entry and $0$ everywhere else, we get that the regret for a certain sequence $\vec{v}$ can be written as:

\begin{equation*}
    \label{loo_multinomial_appendix}
    R_{loo} \left(\vec{v}, q(\cdot)  \right) = \sum_{j=0}^{m-1} \frac{\vec{v}(j)}{N} \log \left( \frac{\frac{\vec{v}(j)}{N}}{q_{\vec{v} - \hat{i}_j}(j)} \right).
\end{equation*}

Thus, one can see that $q_{\vec{e}}(j)$ appears in the regret of a single sequence, $\vec{e} + \hat{i}_j$.
In addition, note that $\vec{v}_k - \hat{i}_0 \in E_k$, while $\vec{v}_k - \hat{i}_j \in E_{k-1}$ for $j>0$, and  that $|V_k| = |E_k|$.

\textbf{General Idea}: Suppose that $R^* < \frac{(m-1)}{N} = \frac{(m-1)(1-\epsilon_0)}{N}$ for some $\epsilon_0 > 0$. We will show by induction that
\begin{equation*}
    \frac{1}{|E_k|} \sum_{\vec{e_k}}q_{\vec{e_k}}(0) \geq 1 - \frac{(m - 1 + k)(1 - \epsilon_k)}{N} + o(\frac{1}{N})
\end{equation*}
where $\epsilon_k \geq 1$ at some point implying that for at least one $\vec{e}\in E_k$, $q_{\vec{e}}(0) > 1$, a contradiction.

To start, for the sequence containing only zeros, represented by $\vec{v}_0$ the regret is:

\begin{equation*}
    R_{\vec{v}_0} = -\log (q_{e_0}(0)) \leq \frac{(m-1)(1-\epsilon_0)}{N}
\end{equation*}
where $e_0(0) = (N-1) \cdot \hat{i}_0$. This leads to:

\begin{equation*}
    q_{e_0}(0) \geq e^{-R}  = 1 - \frac{(m-1)(1-\epsilon_0)}{N} + o(\frac{1}{N})
\end{equation*}

This is the first step of the induction for $k=0$. To make things clearer, we also show explicitly the next step for $k=1$:

{\bf Induction step for $k=1$:}
Since $\sum_{j=0}^{m-1}q_{\vec{e}_0}(i) = 1$, we have the following bound:

\begin{equation}
    \label{multinomial_0_constraint}
    \sum_{j>0}^{m-1} q_{e_0}(0) < \frac{(m-1)(1-\epsilon_0)}{N} + o(\frac{1}{N}).
\end{equation}
Next, consider a sequence with $N-1$ zeros and a single appearance of a letter $j>0$. The regret associated with such a sequence, which we denote by $R_{v_0 - \hat{i}_0 + \hat{i}_j}$ is:

\begin{align*}
    R_{v_0 - \hat{i}_0 + \hat{i}_j} &= \frac{N-1}{N} \log \left( \frac{\frac{N-1}{N}}{q_{e_0 - \hat{i}_0 + \hat{i}_j}(0)} \right) +
    \frac{1}{N}\log \left( \frac{\frac{1}{N}}{q_{e_{(0)}}(j)} \right) \leq \frac{(m-1)(1-\epsilon_0)}{N}.
\end{align*}

where $e_{1, j}$ is the observations vector representing a sequence containing $N-2$ zeros and a single letter $j>0$. Averaging over all $j \in \{1, 2, ..., m-1 \}$ we get:

\begin{align*}
    &\frac{1}{m-1}\sum_{j = 1}^{m-1} \left[ \frac{N-1}{N} \log \left( \frac{\frac{N-1}{N}}{q_{e_{(1,j)}}(0)} \right) +
    \frac{1}{N}\log \left( \frac{\frac{1}{N}}{q_{e_{(0)}}(j)} \right) \right] \leq \frac{(m-1)(1-\epsilon_0)}{N}.
\end{align*}
We now bound from below $\frac{1}{N} \sum_{j=1}^{m-1}\log \left( \frac{\frac{1}{N}}{q_{e_{(0)}}(j)} \right)$. To this end, we will use the constraint \eqref{multinomial_0_constraint}, and form the following Lagrangian:

\begin{align*}
    L = \sum_{j=1}^{m-1}\log(q_{e_0}(j))  + \lambda \cdot\sum_{j=1}^{m-1} q_{e_0}(j) 
\end{align*}
Since every partial derivative with respect to $(q_{e_0}(j))$ is zero, we get:

\begin{equation*}
    \frac{\partial L}{ \partial q_{e_{0}}(j)} = \frac{1}{q_{e_0}(j)}  + \lambda = 0 ,
\end{equation*}
implying that $(q_{e_{(0)}}(j))$ is a constant. Since the total sum is bounded by $\frac{(m-1)(1-\epsilon_0)}{N}$, we get $(q_{e_{(0)}}(j)) = \frac{1-\epsilon_0}{N}$. This leads to:

\begin{equation*}
    \frac{1}{N \cdot (m-1)} \sum_{j=1}^{m-1}\log \left( \frac{\frac{1}{N}}{q_{e_{(0)}}(j)} \right) \geq -\frac{1}{N} \log(1-\epsilon_0)
\end{equation*}
and thus we can get the following upper bound:

\begin{align*}
    &\frac{1}{m-1}\sum \log(q_{e_{(1,j)}}(0)) 
    \\
    &\geq 
    -\frac{(m-1)(1-\epsilon_0) -N \log(\frac{N-1}{N}) + \log(1-\epsilon_0)}{N-1}
    \\
    &= 
    -\frac{m(1-\epsilon_0) + \epsilon_0 + \log(1-\epsilon_0)}{N-1} + o(\frac{1}{N})  
    \\
    &\geq
    -\frac{m(1-\epsilon_1) + \epsilon_0 - \epsilon_0 - \frac{\epsilon^2_0}{2}}{N} + o(\frac{1}{N})  
    \\
    &= 
    -\frac{m(1-\epsilon_1)}{N} + o(\frac{1}{N})
\end{align*}
where $\epsilon_1 = \epsilon_0 + \frac{\epsilon_0^2}{2m}$.
Last, using Jensen's inequality, we get:

\begin{equation*}
    \frac{1}{m-1} \sum_{j=1}^{m-1} q_{e_{(1,j)}}(0) \geq e^{-\frac{m(1-\epsilon_1)}{N}} = 1 - \frac{m(1-\epsilon_1)}{N} + o \left( \frac{1}{N} \right)
\end{equation*}

{\bf The general induction step:}

Assume that we have shown that for some $k-1$, the following holds:

\begin{equation*}
\frac{1}{|E_{k-1}|} \sum_{\vec{e}_{k-1}}q_{e_{k-1}}(0) \geq 1 - \frac{(m + k - 2)(1 - \epsilon_{k-1})}{N}.
\end{equation*}
This immediately implies:

\begin{equation}
\label{induction_constraint}
\frac{1}{|E_{k-1}|} \sum_{j=1}^{m-1}\sum_{\vec{e_{k-1}}}q_{e_{k-1}}(j) \leq \frac{(m + k - 2)(1 - \epsilon_{k-1})}{N}.
\end{equation}

Now, for $k$, consider the average regret over all $v_k \in V_k$, which is naturally upper-bounded by $R^* = \frac{(m-1)(1-\epsilon_0)}{N}$:

\begin{align*}
    &\sum_{v_k} R(\vec{v}_k, q) 
    =\sum_{\vec{v}_k}\left[\sum_{j=0}^{m-1}
    \frac{v_{k}(j)}{N}\log \left( \frac{\frac{\vec{v}_k (j)}{N}}{q_{\vec{v}_k -\hat{i}_j}(j)} \right) \right] 
    \leq  {|V_k|} \cdot \frac{(m-1)(1-\epsilon_0)}{N}
\end{align*}
where by definition $\vec{v}_k(0) = k$. To analyze this expression, we look for the minimal value of the second sum. To this end we incorporate the constraint in \eqref{induction_constraint} and get the Lagrangian:

\begin{align*}
     L &= \frac{1}{|V_k|}\sum_{\vec{v}_k}\sum_{j=1}^{m-1}
    \frac{v_{k}(j)}{N}\log \left( \frac{\frac{\vec{v}(j)}{N}}{q_{\vec{v}_k -\hat{i}_j}(j)} \right)  + \lambda \cdot\frac{1}{|E_{k-1}|} \sum_{j=1}^{m-1}\sum_{\vec{e_{k-1}}}q_{e_{k-1}}(j) 
    \\
    &= \frac{1}{|V_k|}\sum_{\vec{v}_k}\sum_{j=1}^{m-1}
    \frac{v_{k}(j)}{N}\log \left( \frac{\frac{\vec{v}(j)}{N}}{q_{\vec{v}_k -\hat{i}_j}(j)} \right)  
    + \lambda \cdot\frac{1}{|E_{k-1}|} \sum_{j=1}^{m-1}\sum_{\vec{e_{k-1}}}q_{\vec{v}_k - \hat{i}_j}(j)
\end{align*}
By finding the critical point of this Lagrangian, we get:
\begin{equation*}
    q_{\vec{v}_k - \hat{i}_j}(j) = \frac{v_k(j)(1-\epsilon_{k-1})}{N}.
\end{equation*} 
Thus,

\begin{align*}
    &\frac{1}{|V_k|}\sum_{\vec{v}_k}\left[ \frac{N-k}{N} \log \left( \frac{\frac{N-k}{N}}{q_{\vec{v}_k + \hat{i}_0}(0)} \right) \right] 
    \leq \frac{(m-1)(1-\epsilon_0)}{N} -\frac{k(\log(1-\epsilon_{k-1})}{N}.
\end{align*}
which leads to:
\begin{align*}
    &-\frac{1}{|V_k|}\sum_{\vec{v}_k} \log \left({q_{\vec{v}_k + \hat{i}_0}(0)} \right) 
    \\
    &\leq 
    \frac{(m-1)(1-\epsilon_0)}{N} + \frac{k(\log(1-\epsilon_{k-1})}{N} 
    \\
    &- \log \left(1 - \frac{k}{N} \right) + o(\frac{1}{N}) 
    \\ 
    &\leq     \frac{(m-1)(1-\epsilon_0)}{N} - \frac{k(\epsilon_{k-1} + \frac{\epsilon^2_{k-1}}{2})}{N} + \frac{k}{N} + o(\frac{1}{N}) 
    \\
    &= \frac{(m + k - 1)(1 - \epsilon_k)}{N}  + o(\frac{1}{N})
\end{align*}
where:
\begin{equation}
    \label{epsilon}
    \epsilon_k = \epsilon_0 + k \cdot \frac{\epsilon_{k-1} - \epsilon_0 + \frac{\epsilon^2_{k-1}}{2}}{m+k-1}.
\end{equation} 

Last, using Jensen's inequality and the approximation $e^{-\epsilon} = 1-\epsilon + o(\epsilon)$, we get:

\begin{align*}
    \frac{1}{|V_k|}\sum_{\vec{v}_k} {q_{\vec{v}_k + \hat{i}_0}(0)} 
    &\geq 1 - \frac{(m + k - 1)(1 - \epsilon_k)}{N},
\end{align*}
which concludes the induction. 

\vspace{5mm}

To conclude the proof, however, we now show that $\epsilon_k \geq 1$ at some point, in contradiction. To this end, note that by $\eqref{epsilon}$ if $\epsilon_0 > 0$ then $\epsilon_k > \epsilon_0$ for all $k \in \mathbf{N}$. 
In addition,

\begin{equation*}
    \epsilon_{k} - \epsilon_{k-1} = \frac{(m-1)(\epsilon_0 - \epsilon_{k-1}) + k \cdot \frac{\epsilon^2_{k-1}}{2}}{m+k-1} 
    \to \frac{\epsilon^2_{k-1}}{2} > \frac{\epsilon_0^2}{2}
\end{equation*}
This means that at some point $\epsilon_k$ is monotonically increasing, since the the difference between two successive entries is a positive quantity bounded away from zero (actually the difference is also increasing). Thus the sequence diverges and so at some point $\epsilon_k \geq 1$, in contrast to the assumption that $\epsilon_k < 1$. This concludes the proof.

\end{IEEEproof}


\begin{IEEEproof}[Proof of Theorem \ref{theorem_add_1_regret}]
Denote by $\theta_j = \frac{\vec{v}(j)}{N}$ the fraction of appearances of the letter $j$ in the whole sequence. Note that the regret for a specific sequence characterized by $\vec{v}$ is given by:

\begin{equation*}
    \label{loo_multinomial}
    R_{loo} \left(\vec{v}, q(\cdot)  \right) = \sum_{j=0}^{m-1} \theta_j \log \left( \frac{\theta_j}{q_{\vec{v} - \hat{i}_j}(j)} \right).
\end{equation*}
In addition:

\begin{equation*}
    \hat{\theta}_j = \frac{\vec{v}_j - 1}{N - 1} = \theta_j - \frac{1 - \theta_j}{N-1}
\end{equation*}
Thus, we have:

\begin{align*}
    &q_{\vec{v} - \hat{i}_j}(j) =\hat{\theta}_j + \frac{1}{N-1} \left( 1 -m\hat{\theta}_j \right) = 
    \theta_j + \frac{-(1 - \theta_j) + (1  -m \theta_j)}{N-1} + \frac{m(1-\theta_j)}{(N-1)^2} =
    \theta_j - \theta_j \frac{m-1}{N-1} + \frac{m(1-\theta_j)}{(N-1)^2}
\end{align*}
Thus, the associated regret for $q_{\vec{v} - \hat{i}_j}(j)=q_{\vec{e}}(j)$ is
\begin{align*}
    \label{loo_multinomial}
    R_{loo} (\vec{v}, q_{\vec{e}}(j) ) &= -\sum_{j=0}^{m-1} \theta_j \log \left( 1 - \frac{m-1}{N-1} + \frac{m(1-\theta_j)}{\theta_j(N-1)^2} \right)
    \\
    &= \sum_{j=0}^{m-1} \theta_j \left( \frac{m-1}{N-1} + \frac{(m-1)^2}{2(N-1)^2} - \frac{m(1-\theta_j)}{\theta_j(N-1)^2} + o(\frac{1}{N^2}) \right)
    \\
    &= \frac{m-1}{N-1} + \frac{(m-1)^2}{2(N-1)^2} - \frac{m(m - 1)}{(N-1)^2} + o(\frac{1}{N^2}) 
    \\
    &= \frac{m-1}{N} + \frac{m-1 +\frac{1}{2}(m-1)^2 - m(m-1)}{(N-1)^2} + o(\frac{1}{N^2})
    \\
    &= \frac{m-1}{M} + \frac{(m-1)^2}{2(N-1)^2} + o(\frac{1}{N^2})
\end{align*}

\end{IEEEproof}

\begin{IEEEproof}[Proof of Theorem \ref{theorem_equalizer_deterministic_vc_dimension}]
    The proof is very similar to that of Theorem \ref{theorem_multinomial_equalizer}. Assume that the one-inclusion graph of a projection of some hypothesis class $\Theta$ on $x^N$ is a connected graph. Since $y \in \{0, 1 \}$, every edge connects exactly two nodes.

    To prove that the min-max optimal solution $q(\cdot|\cdot)$ must be an equalizer, i.e., the regret it attains is equal for all realizable sequences $y^N$, we will show that one can improve the maximal regret of any solution that is not an equalizer
    
    To this end, first consider a solution $q(\cdot|\cdot)$ that attains its maximal regret over realizable outcome sequences at a specific, single node, which we will denote $N_{max}$. Since the one-inclusion graph is connected, this node must be connected to another node, which we will denote by $N_{connected}$ whose regret is strictly lower. Thus, there must be some $y^N, t$, representing the edge connecting the two nodes, such that the regret of $N_{max}$ is monotonically decreasing with $q(y_t|z^{N \setminus t}, x_t)$ while the regret of $N_{connected}$ is increasing with $q(y_t|z^{N \setminus t}, x_t)$. Thus, and since the logarithmic loss function is continuous and unbounded on $(0,1 )$ we can increase $q(y_t|z^{N \setminus t}, x_t)$ such that the regret of $N_{max}$ will decrease but the regret of $N_{connected}$ will still be strictly lower than the previous value of the maximal regret. Thus, we came up with another probability assignment, which is strictly better in the min-max sense.

    Now assume that there is some $q(\cdot|\cdot)$ that attains its maximal regret over realizable outcomes sequences at several realizable sequences $y^N$, but not at all of them. Again, since the one-inclusion graph is connected, we can find a pair of connected nodes where one of them attains the maximal regret and the other attains a strictly lower regret. We can again alter the probability defined over the edge that connects them and attain a new probability assignment that attains its maximal regret at a strictly lower number of points. We can repeat this process again and again until we again get that the maximal value is attained at a specific node, and by using the process described in the previous paragraph get an altered probability assignment whose maximal regret is strictly lower than the original one.
    
\end{IEEEproof}

\begin{IEEEproof}[Proof of Theorem \ref{theorem_max_min_degree_regret_bound}]
    To prove this upper bound, we will construct a probability assignment and analyze its regret. Naturally, the regret of any specific probability assignment upper bounds the min-max optimal regret.
    
    The probability assignment will be constructed as follows: First, given a graph, we define the outer layer of the graph as the layer that contains all nodes whose degree is at most $k$.

    Next, we shall assign probabilities to all the edges that are connected to the outer layer as follows: If both of the connected edges are in the outer layer the probability assignment will be $\frac{1}{2}$. If the other node, however, is not in the outer layer, we will assign a probability of $\frac{1}{N}$ to the node in the outer layer and $1-\frac{1}{N}$ to the other node.

    After assigning the probabilities for all edges connected to the outer layer, we remove all the nodes in the outer layer and the connected edges and repeat the same process until no more nodes are left.

    Note that since the minimal degree of any sub-graph of the 1-inclusion graph is at most $k$, there will always be nodes in the outer layer, and thus, we are guaranteed that we will deal with all nodes - and thus, assign probabilities to all edges.

    To evaluate the regret associated with this probability assignment, consider a specific node. There are three possible types of edges associated to this node: 
    
    The first type is edges that were connected to this node when it was not in the outer layer. The number of such nodes is only bounded by $N$, but since we assigned a probability of $1-\frac{1}{N}$ for these nodes the associated regret will be bounded by $\frac{N \log(1-\frac{1}{N}}{N}) \simeq \frac{1}{N}$ for large enough values of $N$.

    The second type of edges are those that were connected to this node when both were in the outer layer. The number of such edges is bounded by $k$, and since the probability assignment is $\frac{1}{2}$ the associated regret is bounded $\frac{k \log(2)}{N}$.

    The third part, which will dominate the regret, is edges that connected the node when it was in the outer layer to nodes that belonged to a deeper layer. The number of these edges is again bounded by $k$, but since the probability assignment was $\frac{1}{N}$, the associated regret is bounded by $\frac{k \log(N)}{N}$.

    The regret of each possible edge is also summarized in table \ref{table1}.

\begin{table}[]
\centering
\label{table1}
\caption{Regret Analysis for Theorem \ref{theorem_max_min_degree_regret_bound}}
\begin{tabular}{|l|l|l|l|}
\hline
Connected node in &  Probability  & Number of edges & Maximal loss \\ \hline
Outer Layer               & $1 - \frac{1}{N}$ &  $\leq N$                       & $ \frac{1}{N} + o(\frac{1}{N})$                             \\ \hline
Same Layer                & $\frac{1}{2}$    & $\leq k$                      & $\frac{k \log(2)}{N}$    \\ \hline
Inner Layer               & $\frac{1}{N}$    & $\leq k$                      & $\frac{k \log(N)}{N}$    \\ \hline
\end{tabular}
\end{table}

    Taking all three possible types of edges into account we get a regret which is bounded by $\frac{k \log(N)}{N} + o(\frac{\log(N)}{N})$.

\end{IEEEproof}


\begin{IEEEproof}[Proof of Corollary \ref{corollary_vc_dimension}
]
    Since the VC-dimension is $d$, the density of every sub-graph of the one-inclusion graph is upper-bounded by $d$, see \cite{predicting_01_on_randomly_drawn_points}. Thus, the minimal degree of any sub-graph of the one inclusion graph is $2d$, and the rest follows immediately from \ref{theorem_max_min_degree_regret_bound}.
\end{IEEEproof}


\begin{IEEEproof}[Proof of Theorem \ref{theorem_max_degree_regret_bound}]
Consider a probability Assignment of $\frac{1}{2}$ to every edge. The regret of that probability assignment for every node $N_j$ is just $R_{loo} = \frac{degree(N_j) \log(2)}{N} \leq \frac{ k \log(2)}{N}$.      
\end{IEEEproof}


\begin{IEEEproof}[Proof of Theorem \ref{theorem_d-unique_values_regret}]

    Fix some $d$, and consider a case where $x \in \mathbf{R}$, and the hypothesis class is the "$d$ unique values" hypothesis class, where every hypothesis determines $y$ to be $0$ for all $x$'s except for $d$ specific values, $\theta_1, \theta_2, ... \theta_d$. Each hypothesis can thus be described by $\theta_1, ... \theta_d$. We will show that the regret of this specific hypothesis class is lower bounded by $\frac{d \log(N)}{N} + o \left( \frac{\log(N)}{N} \right)$

 Denote by $R_i$ the regret associated with a sequence containing $i$ ones and by $q_i$ the probability assigned to the next symbol being $1$, given that $i$ ones have been observed. Note that:

\begin{equation*}
    R_i = -\frac{i}{N}\log(q_{i-1}) - \frac{N-i}{N} \log(1 - q_i)
\end{equation*}

    with $R_0 = -\log(1-q_0)$ and $R_d = -\frac{d}{N}\log(q_{d-1})$.
    
    Now, assume that the min-max optimal regret is $R^* = \frac{a \log(N)}{N} - o(\frac{\log(N)}{N})$ is achieved for some $0 \leq a < d $. Since $R_0 \leq R^*$, we get: 
    
    \begin{equation*}
        q_0 \leq 1 - e^{-R^*} = \frac{a\log(N)}{N} + o(\frac{\log(N)}{N}).
    \end{equation*} 
    
    Since $R_1$ is also bounded by $R^*$, we can also lower bound $q_1$ in a similar manner:
    \begin{equation*}
        q_1 \leq 1 - e^{-\left(\frac{(a-1) log(N)}{N - 1} + o \\frac{\log(N)}{N})\right)} < 1 - e^{-\left(\frac{(a-1) log(N)}{N} + o(\frac{\log(N)}{N})\right)}.
    \end{equation*}

We can next use the bound over $q_1$ to get a bound over $q_2$, and so on. We get that for each $k < d$ the following holds:

    \begin{equation*}
        q_k \leq 1 - e^{-\left(\frac{(a-k) log(N)}{N} + o(\frac{\log(N)}{N})\right)} = \frac{(a-k) \log(N)}{N} + o(\frac{\log(N)}{N}).
    \end{equation*}

For $R_d$ we get:
\begin{align*}
R_d 
&= -\frac{d}{N} \log(q_{d-1}) 
\\
&\geq -\frac{d}{n} \log \left( \frac{(a-d-1) \log(N)}{N} + o \left( \frac{\log(N)}{N} \right) \right) 
\\
&= \frac{d \log(N)}{N} + o \left( \frac{\log(N)}{N} \right) 
\\
&> \frac{a \log(N)}{N} o \left( \frac{\log(N)}{N} \right)
\end{align*}

and thus we have a sequence for which the regret is larger than $\frac{a\log(N)}{N}$, in contradiction to the assumption that $R^* \leq \frac{a \log(N)}{N} + o \left( \frac{\log(N)}{N} \right)$

\end{IEEEproof}
    

\begin{IEEEproof}[Proof of Theorem \ref{theorem_multiclass}]

\textbf{General Idea:} The proof will be quite similar to that of Theorem \ref{theorem_max_min_degree_regret_bound}, with a slight complication because we are now dealing with a hyper-graph. We will again define an outer layer, assign all probabilities connected to it, remove the nodes in the outer layer, and repeat; yet, because we are dealing with a hyper-graph, we will, in some cases, assign probabilities to edges over several layers.

Here, we shall define the outer layer as the layer that contains all nodes whose degree is at most $\mu$. Now, consider an edge connected to a node in the outer layer, and assume that we already assigned $ 0 \leq p < 1$ of the total probabilities of this edge to previously dealt-with nodes. Now, if all the remaining nodes connected in this edge are in the same layer, we will distribute the remaining probability $p$ equally, and say that we "closed" the edge in this layer. Otherwise, if the edge contains nodes in an inner layer, we will assign a probability of $\frac{1}{(m-1) \cdot N}$ to each of the nodes currently in the outer layer. We can now remove all the nodes in the outer layer, and proceed in the same manner with the new outer layer.

Before we analyze the obtained regret, note that since the number of nodes connected by some edge is bounded by the alphabet size of the outcomes space, $m$. Thus, it is easy to verify that when we have an edge connected to a node in the outer layer, the sum of all previously assigned probabilities in this edge is no more than $(m-1)\frac{1}{(m-1) \cdot N} = \frac{1}{N}$. In addition, the minimal probability assigned to some node in a specific edge is $\frac{1}{(m-1) \cdot N}$.

We can now show an upper bound over the regret attained by our probability assignment for any node: 

First, note that for edges that were "closed" in the layer to which the node belongs, and did not contain another node from the same layer, the probability assigned to the node in this edge is at least $1-\frac{1}{N}$, and thus the contribution of those edges to the regret is at most $-\log \left( 1- \frac{1}{N} \right) = \frac{1}{N} + o \left( \frac{1}{N} \right)$. 

Next, note that the maximal number of edges that contained nodes either from the same layer or inner layers is $\mu$, the minimal probabilities assigned to the node in these edges is $\frac{1}{(m-1) \cdot N}$, and thus we get can bound the contribution of those edges to the regret of the node by $\frac{\mu \log(N) + \log(m-1)}{N}$. 

Combining these two bounds we get that the total regret for every node is upper-bounded by $\frac{\mu \log(N)}{N} + o \left( \frac{1}{N} \right)$.

\end{IEEEproof}


\begin{IEEEproof}[Proof of Theorem \ref{theorem_probabilistic_vc_dimension}]
    \textbf{General Idea}: The proof consists of constructing from the possible partitions of the data features $x$-s into groups a set of experts, and utilizing known results regarding learning with experts advice.
    
    First, note that the learner knows the whole sequence $x^N$, no matter which $x_t$ is the test, and thus knows all the possible partitions of the entire sequence no matter what the test data feature is. Building upon this observation, we will construct a probability assignment using a mixture of exponentially weighted experts as follows: We define each possible partition of $x^N$ into groups as a separate expert. Now, given some test data feature $x_t$, each expert will assign a probability for $y_t$ based on Laplace's rule of succession for an alphabet of size $m$. Thus, the $\ell$-th expert will assign a probability:

    \begin{equation*}
        p_{\ell}(y_t = i|x^N, y^{N \setminus t}) = \frac{n_j(i) + 1}{N_j  + m} 
    \end{equation*}
    where $j$ is the group to which $x_t$ is assigned by the partition associated with this $\ell$-th expert, $n_{j}(i)$ is the number of appearances of the letter $i$ in the $j$-th group excluding $y_t$ and $N_{j}$ is the total number of data features assigned to the $j$-th group excluding $y_t$.  
    
    To utilize known results from the field of learning with experts' advice, we will take advantage of the fact that the $x$-s are different (which is common when the feature space is large) and define some arbitrary order for them. Then, the weight of the expert when the test's feature is $x_t$ will be proportional to the exponent of his cumulative loss on $z^{t-1}$. Thus, the weight $w_\ell(t)$ assigned to the $\ell$-th expert in time $t$ will be:

    \begin{align*}
        w_\ell(t) = \frac{\exp{\sum_{i=1}^{t-1}\log(p_\ell(y_i|x^N, y^{t-1})}}{\sum_{\ell' \in G(\Theta, x^N)}\exp{\sum_{i=1}^{t-1}\log(p_{\ell'}(y_i|x^N, y^{t-1})}}
    \end{align*}
    where $G(\Theta, x^N)$ is the set of partitions of $x^N$ into different groups induced by $\Theta$, that defines our experts. Note that while the experts depend upon all the outcomes except for the one on which they are tested, their weights are only a function of their past performance in a sequential manner. Thus, we can invoke Theorem 3.2 and Proposition 3.1 in \cite{cesa2006prediction} and get that the difference in normalized cumulative losses between the learner and the best expert in hindsight is bounded by $\frac{|G(\Theta, x^N)|}{N}$ 

    

    Utilizing Sauer–Shelah lemma, which bounds the number of these experts by $(eN/d)^d$, the resulting normalized difference between the cumulative losses is bounded by:
    
    \begin{align*}
        \frac{d\log(N)}{N} + O(\frac{1}{N}).    
    \end{align*}

    Next, we can utilize corollary 3, to bound the regret of the said expert. To this end, we will look at each group separately and note that for each group, the regret with respect to the best probability assignment behaves as $\frac{(m-1)}{N_j} + o\left(\frac{1}{N_j}\right)$. All that is left is to weigh those regrets according to the number of samples on which they appear to get that the difference between the normalized cumulative losses is:
    \begin{equation*}
        \frac{N_0}{N} \frac{(m-1)}{N_0} + \frac{N_1}{N} \frac{(m-1)}{N_1} + o \left( \frac{1}{N}\right) = \frac{(m-1)}{N} + o \left( \frac{1}{N}\right) 
    \end{equation*}
  
    Combining both, the maximal regret of this learner is still upper bounded  $\frac{d \log(N)}{N} + O(\frac{1}{N})$.                   
\end{IEEEproof}

\begin{IEEEproof}[Proof of Theorem \ref{theorem_pNMLs_fail}] 

Note that when the training set contains only $0$, then for $\theta=x_N$ we get $p_{\theta})(y^N|x^N) = p_{\theta})(y_N|x_N) = 1$ for $y=1$, while for $\theta: \forall_{i \in \left[1, ..., N \right]} \theta \neq x_i$, we get  $p_{\theta})(y^N|x^N) = p_{\theta})(y_N|x_N) = 1$ for $y=0$. Thus, both the pNML and pNML-2 are $\frac{1}{2}$ and the regret is $\log(2)$ regardless of $N$.

\end{IEEEproof}
\end{appendices}
\bibliographystyle{IEEEtran}
\bibliography{references}
\end{document}